# Accounting for Uncertainty During a Pandemic

Jon Zelner, Julien Riou, Ruth Etzioni, and Andrew Gelman

12 June 2020

*Abstract.* We discuss several issues of statistical design, data collection, analysis, communication, and decision making that have arisen in recent and ongoing coronavirus studies, focusing on tools for assessment and propagation of uncertainty. This paper does not purport to be a comprehensive survey of the research literature; rather, we use examples to illustrate statistical points that we think are important.

## 1. Statistics and uncertainty

Just as war makes every citizen into an amateur geographer and tactician, a pandemic makes epidemiologists of us all. Instead of maps with colored pins, we have charts of exposure and death counts; people on the street argue about infection fatality rates and herd immunity the way that in the past they might have debated strategies and alliances.

The severe acute respiratory syndrome coronavirus 2 (SARS-CoV-2) pandemic has brought statistics and uncertainty assessment to public discourse to an extent rarely seen except in election season and the occasional billion-dollar lottery jackpot. Statistical claims become political claims and vice-versa. As statisticians and epidemiologists, we can try to contribute to the discourse by laying out some of the challenges that arise in assessing uncertainty and propagating it through statistical analysis and decision making. We consider several examples and conclude with some general recommendations.

Statistics is key throughout the life cycle of a scientific project, from design through data collection and analysis, and ultimately through communication of results for decision recommendations. In the case of a pandemic like SARS-CoV-2, surveillance data are critical for assessment of current status and for future projection, and clinical measurements are vital for evaluating diagnostic tests and intervention efficacy. Design includes sample-size calculations, determination of comparison groups, and time horizons, and randomization,and is critical in research to identify effective treatments and vaccines. Analysis includes evaluation and estimation based on clinical studies as well as disease modeling studies for forecasting and decision support. Communication includes the challenge of drawing inferences and making decisions based on a variety of models and data sources. Uncertainty is present at each step.

## 2. Data and measurement quality

It is becoming painfully apparent that the numbers defining the global burden of SARS-CoV-2 are at best uncertain and at worst frequently wrong. The bread and butter of disease

surveillance—cases and deaths—are both suspect, for reasons that are only beginning to be fully understood. Studies that rely on these as inputs, for example for estimating transmission dynamics or case fatality rates, have commonly made the mistake of considering these numbers as given (and reliable) and do not account for uncertainty or bias in reporting.

There has been some compelling reporting on how the number of deaths reported in the first few months of the pandemic far exceeds what would have been expected at this time of the year, particularly in states like New York.  There has also been good reporting how the differences across states in reporting of covid-related deaths may be complicating things (Harmon, 2020).  Some states have changed how they classify a death as due to the virus, leading to potential increases in death counts in some cases (e.g. Michigan) and reductions in death counts (e.g. Colorado).  One big question is getting at how the changes in test availability and distribution both between regions and groups, and over time (for example, as a result of inadequate infrastructure and reagent shortages), impacts our assessment of incidence, prevalence, and mortality, conditional on age and other demographic variables.  Larremore et al. (2020) deals with some of the questions about measuring seroprevalence but doesn't engage with any of the pesky issues surrounding social bias in testing.

One way to address data quality is to triangulate.  In a clinical study, a hospital can perform antibody tests and RT-PCR RNA tests on patients.  In a study tracking symptoms, data can be collected from multiple sources, as in the Carnegie Mellon project that tracks Facebook and Google surveys, hospital records, web searches and flu tests (Rosenfeld et al., 2020).

When measurements cannot be easily calibrated, inferences can be sensitive to assumptions; for example, the controversial Stanford antibody study (Bendavid et al., 2020, Gelman and Carpenter, 2020) was sensitive to the assumed sensitivity and specificity of the antibody test.

These issues are no less pronounced when contemplating population-level transmission dynamics. The basic reproduction number, $R_0$, and its cousin the effective reproduction number, $R$, which measures the actual number of infections generated by an average case, are often cited as measures of epidemic control.. However, it is easy to forget that $R_0$ and $R$ are not empirical quantities. They are estimated on the basis of surveillance data, which as noted above, is not as reliable as we might wish to believe. In addition, $R$ is a function of (a) the per-contact infectiousness of each individual, and (b) the rate at which those contacts occur. Reduce either or both of these and you are likely to reduce the rate of spread.  Both are subject to between-individual variation, due, for example, to variable compliance with social distancing efforts, variation in the extent of viral shedding, age specific differences in contact and infectiousness. This variation is widely understood in infectious disease epidemiology, and there are theoretical and statistical modeling frameworks that allow us to account for inter-individual variability in susceptibility and infectiousness.

Drivers of variation in infectiousness and susceptibility at an individual or population level can be studied using a hierarchical approach.  In this area, there are at least three key dimensions of uncertainty that we need to consider: (1) What range of values of the average infectiousness is consistent with the observed data? (2) How much between-individual variation is there in infectiousness/susceptibility, and how much does it matter to address it specifically? (3) If we implement an intervention to reduce the value of $R_0$, how can we estimate how well it worked? The Imperial College group has fit some reasonable models (for example, Unwin et al., 2020) trying to untangle effects of different policies on the spread of coronavirus, making use of variation in space and time of the growth rates of the infection,

### 3. Design of clinical trials for treatments and vaccines

Part of designing a study is accounting for uncertainty in effect sizes. Unfortunately there is a tradition in clinical trials of making optimistic assumptions in order to claim high power. Here is an example that came up in March, 2020. A doctor was designing a trial for an existing drug that he thought could be effective for high-risk coronavirus patients. He contacted one of us to check his sample size calculation: under the assumption that the drug increased survival rate by 25 percentage points, a sample size of N = 126 would assure 80% power.[1] When we asked him how confident he was in his guessed effect size, the doctor replied that he thought the effect on these patients would be higher and that 25 percentage points was a conservative estimate. At the same time, he recognized that the drug might not work. We asked the doctor if he would be interested in increasing his sample size so he could detect a 10 percentage point increase in survival, for example, but he said that this would not be necessary.

It might seem reasonable to suppose that a drug might not be effective but would have a large individual effect in case of success. But this vision of uncertainty has problems. Suppose, for example, that the survival rate was 30% among the patients who do not receive this new drug and 55% among the treatment group. Then in a population of 1000 people, it could be that the drug has no effect on the 300 of people who would live either way, no effect on the 450 who would die either way, and it would save the lives of the remaining 250 patients. There are other possibilities consistent with a 25 percentage point benefit--for example the drug could save 350 people while killing 100--but we will stick with the simple scenario for now. In any case, the point is that the posited benefit of the drug is not "a 25 percentage point benefit" for each patient; rather, it's a benefit on 25% of the patients. And, from that perspective, *of course* the drug could work but only on 10% of the patients. Once we've accepted the idea that the drug works on some people and not others--or in some comorbidity scenarios and not others--we realize that "the treatment effect" in any given study will depend entirely on the patient mix. There is no underlying number representing the effect of the drug. Ideally one would like to know what sorts of patients the treatment would help, but in a clinical trial it is enough to show that there is some clear average effect. Our point is that if we consider the treatment effect in the context of variation between patients, this can be the first step in a more grounded understanding of effect size.

---

[1] With 126 people divided evenly in two groups, the standard error of the difference in proportions is bounded above by √(0.5*0.5/63 + 0.5*0.5/63) = 0.089, so an effect of 0.25 is at least 2.8 standard errors from zero, which is the condition for 80% power for the z-test.

When considering design for a clinical trials more generally, we recommend assigning cost and benefits and balancing the following:

– Benefit (or cost) of possible reduced (or increased) mortality and morbidity from COVID-19 in the trial itself.
– Cost of toxicity or side effects in the trial itself.
– Public health benefits of learning that the therapy works, as soon as possible.
– Economic / public confidence benefits of learning that the therapy works, as soon as possible.
– Benefits of learning that the therapy doesn't work, as soon as possible, if it really doesn't work.
– Scientific insights gained from intermediate measurements or secondary data analysis.
– Financial cost of the study itself, as well as opportunity cost if it reduces your effort to test something else.

This may look like a mess—but if you're not addressing these issues explicitly, you're addressing them implicitly. Whatever therapies are being tried, should be monitored. Doctors should have some freedom to experiment, and they should be recording what happens. To put it another way, they're trying different therapies anyway, so let's try to get something useful out of all that. It's also not just about "what works" or "does a particular drug work," but how to go about understanding what works, when, and for whom.  You want to get something like optimal dosing, which could depend on individuals.  But you won't get good discrimination on this from a standard clinical trial or set of clinical trials.  So we have to go beyond the learning-from-clinical-trial paradigm, designing large studies that mix experiment and observation to get insight into dosing, subgroup effects, and other practical questions.

Also, lots of the relevant decisions will be made at the system level, not the individual level. These issues of decision making are crucial, and they go beyond the standard clinical-trial paradigm.

Other issues arise when designing clinical trials for vaccines.  Lumley (2020) writes:

> There are over 100 potential vaccines being developed, and several are already in preliminary testing in humans. There are three steps to testing a vaccine: showing that it doesn't have any common, nasty side effects; showing that it raises antibodies; showing that vaccinated people don't get COVID-19.
>
> The last step is the big one, especially if you want it fast. . . . We don't expect perfection, and if a vaccine truly reduces the infection rate by 50% it would be a serious mistake to discard it as useless. But if the control-group infection rate over a couple of months is a high-but-maybe-plausible 0.2% that means 600,000 people in the trial—one of the largest clinical trials in history.

> How can that be reduced? If the trial was done somewhere with out-of-control disease transmission, the rate of infection in controls might be 5% and a moderately large trial would be sufficient. But doing a randomised trial in a setting like that is hard—and ethically dubious if it's a developing-world population that won't be getting a successful vaccine any time soon. If the trial took a couple of years, rather than a couple of months, the infection rate could be 3-4 times lower—but we can't afford to wait a couple of years.
>
> The other possibility is deliberate infection. If you deliberately exposed trial participants to the coronavirus, you could run a trial with only hundreds of participants, and no more COVID deaths, in total, than a larger trial. But signing people up for deliberate exposure to a potentially deadly infection when half of them are getting placebo is something you don't want to do without very careful consideration and widespread consultation. . . .

And Delaney (2020) follows up:

> One major barrier is manufacturing the doses, especially since we decided to off-shore a lot of our biomedical capacity in the name of efficiency (at the cost of robustness). . . . We want an effective vaccine and it may be the case that candidates vary in their effectiveness. There are successful vaccines that do not grant 100% immunity. The original polio vaccines were only 60-70% effective versus one of the strains, but that still led to a vast decrease in the number of infections in the United States once vaccination became standard.
>
> So, clearly we want trials. . . . Now we get to the point about medical ethics. A phase III trial takes a long time to conduct and there is some political pressure for a fast solution. . . . if the virus is mostly under control, you need a lot of people and a long time to evaluate the effectiveness of a vaccine. People are rarely exposed so it takes a long time for differences in cases between the arms to show up. . . .
>
> Another option is the challenge trial. Likely only taking a few hundred participants, it would have no more deaths than a regular trial. But it would involve infecting people, treated with a placebo(!!), with a potentially fatal infectious disease. There are greater good arguments here, but the longer I think about them the more dubious they get to me. Informed consent for things that are so dangerous really does suggest coercion. . . .

It seems hard to imagine that clinical trials for treatments could be organized efficiently across research groups, given the current competitive nature of biomedical research. But maybe it could be possible to organize clinical trials for vaccines. One advantage would be that it should not be necessary to include a placebo arm with each trial.

## 4. Disease models

Infectious disease transmission models have been held to unprecedented and deserved scrutiny during the COVID-19 crisis. The field of infectious disease modelling finds its roots in the work of Ross (1910) on malaria, using mathematical tools to describe the complex relations between parasites, vectors, and hosts. Ross defined the concept of dependent happenings, whereby the frequency of an event such as an infection in an individual depends on the number of individuals already affected (Halloran, 1991). Kermack and McKendrick (1927) formalized this approach, leading to the development of the SIR (susceptible-infectious-recovered) differential equation system that is still the basis of many of the models used for SARS-CoV-2 today. In the SIR model, the processes of contagion and immunity are modeled following the mass action principle: the incidence of new infections is dependent on the proportion of infectious and susceptible individuals in the population, assuming homogeneous mixing. In the following decades, the field of infectious disease modeling has seen tremendous development but has long been kept separated from statistical modeling and inference. The focus was on putting theory into equations and exploring different scenarios, leading to important developments in the development and understanding of interventions aimed at controlling epidemics such as vaccines or vector control. Until recently, comparatively less attention has been given to statistical concepts such as inference, measurement, and uncertainty.

Several types of approaches have been used to model the transmission of SARS-CoV-2, depending on the stage of the epidemic and the objectives of the work.

Whether the objective of a model is inference or forecasting, the handling of uncertainty should remain a constant concern. We can distinguish three sources of uncertainty. Stochastic uncertainty arises from chance events during the course of transmission (whether a contact between an infectious and a susceptible person will result in transmission) or data generation (sampling variation in infected individuals that are reported as cases). Parameter uncertainty represents the imperfect level of knowledge of a particular quantity, such as the average duration of the incubation period which is a fixed input parameter to most transmission models. Model (or structural) uncertainty refers to the set of assumptions underlying any modeling attempt and their adequacy to reality (Draper, 1995). To avoid overconfidence, especially when results are expected to impact policy, one should acknowledge and discuss the potential impact of each of these sources of uncertainty, and as often as possible directly propagate the uncertainty into the results.

In the very early stages of the emergence of SARS-CoV-2 in Wuhan, China, the focus has been put on estimating the basic reproduction number $R_0$ from surveillance data on cases of SARS-CoV-2 infection. $R_0$ is defined as the average number of secondary cases that are generated by an infectious individual in a fully susceptible population. In the first few weeks after its emergence, it was reasonable to assume that the population was fully susceptible to SARS-CoV-2 infection, allowing the use of simple models based on branching processes or exponential growth. Estimating $R_0$ from counts of reported cases constitutes a typical inference problem and must account for important considerations regarding stochastic, parameter, and model uncertainty.

In the context of emerging pathogens, stochastic uncertainty can be very impactful. As very few people are affected, any outlier behaviour may have a strong impact on the course of the disease. One key component here is the assumed distribution in the number of secondary cases. In a totally-susceptible population, its average is by definition $R_0$, but this can vary from individual to individual, with the extreme being a superspreading event (defined as an usually large number of secondary cases generated by a single infectious person). Superspreading events can have a considerable impact in the early stages of disease emergence by accelerating the spatial spread of the pathogen, as was seen for instance during the emergence of Middle-East Respiratory Syndrome coronavirus (Kucharski and Althaus, 2015). Individual heterogeneity and the potential for superspreading events can be accounted for using a negative binomial distribution for modelling the number of secondary cases (Lloyd-Smith et al., 2005).

Estimating $R_0$ using data from the early incidence of SARS-CoV-2 infection in Wuhan also carries the problems of parameter and model uncertainty. Examining the mechanisms leading to the generation of count data gives insight about the basic assumptions that will explicitly or implicitly be part of any modeling attempt: (1) an initial zoonotic event led to the infection of a number of humans on a given date; (2) each of these cases generated secondary cases ($R_0$ cases on average, with a distribution as discussed above); (3) each of these secondary cases generated cases, with a delay that corresponds to the generation time (the gap between two successive generations of cases, which also is a random variable, not a constant); (4) infected cases will have an incubation period, some of the cases will have symptoms, some of the symptomatic cases will seek care, some of the patients will be tested and diagnosed, some of the diagnosed will be reported to the authorities and counted as a case. From these observations, we understand that is not possible to estimate at the same time $R_0$ the date and size of the initial zoonotic event, the incubation period, and the generation time from information about the incidence of SARS-CoV-2, as several combinations of these parameters may lead to the same data. To estimate $R_0$, it is therefore necessary to bring external information about the other parameters. Here enters parameter uncertainty, as overconfidence about the initial conditions or the generation time would result in overconfidence about the value of $R_0$.

Thinking about the mechanisms of data generation brings further considerations about model uncertainty. To this day, very few details are known about the situation surrounding the emergence of SARS-CoV-2 in Wuhan at the end of 2019. Putting aside any political aspect, the early phase of emergence of an unknown pathogen is always a chaotic matter. Explicitly or implicitly, modeling the transmission of SARS-CoV-2 in this context will require strong assumptions about how this data was generated. For instance, some authors took the number of reported cases in Wuhan in the first few weeks at face value and directly inferred the rate of exponential growth and thus $R_0$, implicitly assuming that the proportion of ascertainment (the proportion of cases that end up in the data) was constant over the period considered (Majumder and Mandi, 2020, Li et al., 2020).  Other authors made explicit assumptions about the shape of variation of ascertainment with time (Zhao et al., 2020).  Rather than making assumptions about ascertainment in Wuhan, other authors prefered to use data on national and international cases of SARS-CoV-2 identified in areas still unaffected by the turmoil together with traffic data (Read, 2020, Riou and Althaus, 2020, Imai et al., 2020).  However, this approach carries other assumptions about the representativity of people who traveled from Wuhan to other places.  Differences across estimates based on different assumptions may be referred to as model uncertainty, and is in itself a good reason to consider multiple approaches to study the same issue.

Beyond the first few weeks following emergence, it becomes more and more difficult to continue to assume that transmission continues to take place unhindered. Whether the objective is prediction on inference, it becomes crucial to account for immunity, social distancing or control measures in addition to just contagion.  Two broad categories of transmission models are adapted to this task.  Agent-based models are used to simulate the individual behaviour of agents and can get as intricate as imaginable, going as far as to simulate every vehicle moving in a country (Abhari, Marini, and Chokan, 2020).  These models can provide insight but can be difficult or impossible to fit to data.  In contrast, compartmental models divide the population into different states (e.g. susceptible, infectious, and removed for the classical SIR model), without considering any difference among individuals within a state.  Compartmental models may be considered within a stochastic or a deterministic framework.  The stochastic framework considers the probability of occurrence of each event at each time step, and as hinted by its name is better suited to handle stochastic uncertainty.  The deterministic framework relies upon solving systems of ordinary differential equations (ODE) and leads to the same average results when the number of infected is sufficiently large.  The reduction in computational cost associated with solving ODEs instead of simulating a large number of events is important when the objective is inference, for example when using Bayesian software such as Stan (Stan Development Team, 2020).

In addition to these two approaches a third approach that is more empirically based was developed and publicized by the Institute for Health Metrics and Evaluation (IHME).  The IHME model assumed a Gaussian curve for the shape of the epidemic's mortality trajectory, empirically estimated how restrictions including social distancing enacted in China impacted the time to and height of the peak, and then extrapolated to other settings on the basis of their accumulating mortality data.  The symmetry assumption coupled with the rapid rise in cases and deaths in almost every region, meant that the IHME model predicted a much more rapid decline than other models.  As the virus spread across the U.S. it became clear that the model was not correct and the IHME recently revised the model, replacing it with a hybrid empirical-compartmental approach.

As model complexity increases, the proper handling of uncertainty becomes even more essential.  Model predictions should incorporate stochastic uncertainty by including prediction intervals. Parameter uncertainty should directly be propagated in the results. The quantification of uncertainty in the model outcomes is an integral part of the results and should not be relegated to the side as sensitivity analyses. In this regard, the Bayesian framework with its focus on parameter probability distributions is attractive. Model uncertainty can be handled by carefully considering whether the model structure and all relevant assumptions (even implicit) are adapted to the question as well as using technical tools such as stacking (Liu et al., 2018). Conducting sensitivity analyses with alternative models is always sensible, but there is only so much than a team can do about its own model. It is advisable to rely on other researchers and experts to provide critical assessment of the model by releasing code and data on an appropriate platform. Model uncertainty is best assessed by the community, and this requires transparency. Code sharing will also bring to academia much-needed good practices for programming, and in the long run build more confidence in the field of infectious disease modeling. Additionally, it is important to acknowledge the biases associated with some approaches, as for instance, fitting to cumulative incidence curves is known to lead to bias and overconfidence (King et al., 2015).

### 5. Statistical analysis

So far we have discussed accounting for uncertainty in design, data,, and modeling of epidemics. In addition, data analysis can account for uncertainty and variation using multilevel modeling all the way, and decision making can be based on costs and benefits, not statistical significance. We have relatively little to say about statistical analysis because this is one area in which there are readily-available tools to handle uncertainty and variation.

We are aware of several coronavirus analyses that make use of multilevel models and Bayesian inference. Unwin et al. (2020) is an analysis by the Imperial College group that partially pools across U.S. states, and they have done similar analyses for Europe (Flaxman et al., 2020, Vollmer et al., 2020). Gelman and Carpenter (2020) reanalyze the Santa Clara antibody study using a hierarchical Bayesian model to account for variation of test specificities and sensitivities. A partial list of coronavirus projects using the Bayesian inference engine Stan appears at Stan Forums (2020). Bayesian analysis can also be performed in the data collection stage, allowing more efficient designs (Harrell, 2020).

A challenging issue with statistical models fit during an ongoing epidemic is unobserved/partially observed data. This isn't that much of an issue with deterministic population-level models, but it wreaks havoc on stochastic differential equation models and individual-level approaches. As a result, deterministic models have had wide influence, despite their weaknesses and often in situations where demographic stochasticity of the transmission process should be accounted for, along with the impact of individual-level variation on outcomes.

Early statistical inferences for epidemic models were actually rooted in a stochastic approach known as the TSIR (time-series SIR) model which was originally used to account for time-varying birthrates and demographic stochasticity in models of measles transmission (see

Wakefield et al., 2018), but this has been surpassed by far more complex, far less understandable approaches.  An appealing aspect of the TSIR is that it is just a transformation of a regression model and so is accessible to researchers and policymakers with statistical training.

## 6. Communication

To effectively communicate the results of analyses conducted during the pandemic, what they are meant to accomplish needs to be clear.  In the context of the COVID-19 pandemic, this raises the problem of effective scientific communication to the central place it has always belonged.  This includes communication of key dimensions of uncertainty in risk.  One of the key challenges here is familiar:  how does one impart a gestalt understanding of an interval statistic, such as a confidence or credible interval, to as broad of an audience as possible (van der Bles et al., 2020)?  Another challenge relates to communication of the different ways in which uncertainty arises and the difficulty of picking one apart from another.  For example, what do we do when we can't figure out if we're seeing the results of process noise, observation noise, observation bias, or some combination thereof?

Much of the controversy surrounding the multiple transmission models used for prediction and planning could be mitigated by a more pragmatic reframing of what these—and all mathematical and statistical models—are all about.  Namely, they distill assumptions and data into inferences for outcomes of interest.  Understood this way, they are primarily tools for dimension reduction and exploration, rather than divining rods.  Nonetheless, there are political and emotional reasons people may glom on to one model or another, that make a more measured appreciation of the ability of models to dispel uncomfortable or politically inconvenient uncertainty, unlikely to be broadly achieved any time soon.  Nonetheless, individual scientists have to balance the career and emotional imperatives to deliver impactful, influential results with a more realistic—and less "scientific"—take on what it is that we are doing.

We have also seen some high-priority junk science, such as extreme results reported from small uncontrolled trials, an observational study that may have been based on fake data, and a leaching of trust from formerly respected institutions such as the Center for Disease Control, the World Health Organization, and leading universities and medical journals.  It is good news that bad research can be rapidly debunked on social media, but it is not clear what the new equilibrium will be.  Going beyond problems with individual studies, we recommend more emphasis on accurately communicating uncertainty in model inferences and predictions, as discussed for example by Hullman et al. (2019).

In Etzioni (2020) we explain our discomfort with communication of model-based projections from the IHME:  the issue is not just with the model but also with how any changes in projections based on updating the inputs or the model itself were reported. In the case of the IHME, the model is an empirical model, so its projections kept changing as the model adapted to accumulating data on the epidemic in the United States.  Ultimately, it became clear that key model assumptions rendered it incorrect for making such projections and the model itself was completely revised, changing the projected cumulative deaths considerably.  In and of itself, this demonstrated the uncertainty inherent in the model.  But the new projections were not attributed to the uncertainty or the model.  They were attributed to states beginning to relax social distancing.  This interpretation not only lends itself to politicization, it completely understates the uncertainty in the model, suggesting not only that the updated projection is correct, but that the previous projection was as well.

One thing we keep hearing in conversations with state government officials is a concern that people just don't understand when they are at risk.  Maps and other visuals can  give a realistic and visceral sense of what that risk looks like.  Many questions of science communication arise here that relate specifically to the translation of theory into models and models into spoken and written language.  Meanwhile, pundits and public intellectuals muddy the waters by naively interpreting the psychology literature on risk perception and not taking into account the unpredictability of contagion; see Epstein (2020) and Sunstein (2020) for examples of such punditry and Douglass (2020) and Cirillo and Taleb (2020) for criticisms of shallow individualistic risk analysis.

Another problem relates to the communication of uncertainty in the structure of the models themselves.  We've seen an appetite both from the public and from modelers themselves to find the one true model, with that Box quote (which, like the term "social distancing," we hope never to hear again after this year) tacked on to papers and talks as a kind of fig leaf.  The modeling literature in epidemiology for the large part focuses on validating models through the fit of the curve to the data and demonstrating some accuracy in short-term temporal prediction.  But researchers have traditionally been far less interested in dealing with the distribution of infection in the population.  Hierarchical models for examining both individual level and ecological outcomes are important, as would an in-depth discussion of how to align model validation with public health goals:  should our model maximize short term prediction at the population level or give a better long-term sense of who will be at risk in terms of demographic and geographic characteristics.

### 7.  Information aggregation and decision making

At the time of this writing, there is vigorous debate in the news media, social media, and governments regarding possible future paths of the epidemic and how best to mitigate it.  In the United States, these disputes have taken on a political dimension:  major politicians in both parties have oscillated between complacency and hysteria, and a persistent partisan divide has emerged around both the perception of risk and the use of protective equipment such as masks.  There has also been a shift from an initial phase of transmission in more densely-populated urban and suburban areas, to growth in more rural areas with different issues relating to healthcare access, patterns of contact, and underlying vulnerabilities to infection and severe disease relating to age and comorbidities.  Globally, too, the pandemic has begun to drift southward, with recent surges in transmission in Latin America and South Asia.

Here, though, we focus on the statistical question of combining multiple sources of uncertain information. On the risk side is the news of uncontrolled spread of the virus in early 2020 in Wuhan and then northern Italy, followed by the scary charts showing a doubling every few days during the period when most of the world was conducting business as usual. On the "don't panic" side are the many countries around the world, and many states within the U.S., which still have seen very few coronavirus cases and whose hospitals have not been overwhelmed.

One thing that troubled us in the earliest phases of the pandemic response was the emphasis on rapid analysis of complex, incomplete datasets, followed by rapid publication and extensive media coverage. Rapid response is not inherently problematic, but the conjuring of theoretical frameworks and analytic tools on the fly is unlikely to benefit many more people than the authors of the study. Instead, this makes more sense when you have an existing framework and set of tools that you can apply with minor modifications to incoming data, as was the case with a number of groups enlisted in the earliest days of the pandemic, including IHME as well as Imperial and other groups.

This leads us to wonder whether some kind of disaster model pre-registration is in order for future events, so that the generic behavior of the set of potential tools is well understood before being pressed into services. This could be looser than a clinical trial registration but at least gives the key data inputs and outputs and some characterization of expected behavior under different scenarios. Critically, some type of standardization would give the ability to engineer connections between different types of analyses, so that information on, for example, variable PCR testing across geographic areas and demographic groups, can be easily used to inform estimates of disease incidence and prevalence.

This takes us back to the motivating question behind this essay: How can we adequately account for uncertainty in a pandemic? The question is probably better reframed as: How can we be better *prepared* to address the uncertainty inherent in the response to the next pandemic or another catastrophic, unforeseen—but foreseeable—event. An answer to this question may lie in a reimagining of the tools of epidemiological modeling from something that looks a bit more like the engineering perspective and a bit less like the "pure science" perspective. This entails a move away from analyses as one-off exercises that uncover some permanent—or at least durable—truth, towards a more software-like, continuous-improvement conception of the products of statistical analysis.

**References**

Abhari, R. S., Marini, M., and Chokani, N. (2020). COVID-19 epidemic in Switzerland: Growth prediction and containment strategy using artificial intelligence and big data. https://www.medrxiv.org/content/10.1101/2020.03.30.20047472v2